\documentclass[prd,reprint,superscriptaddress,nofootinbib,amsmath,subcaption]{revtex4-1}
\usepackage[dvips]{graphics}
\usepackage{subfigure}
\usepackage{epstopdf}
\usepackage{lineno,hyperref}
\usepackage{graphicx}
\usepackage{amssymb}
\usepackage{bm}
\usepackage{color}
\usepackage{multirow}
\usepackage{dcolumn}
\usepackage{makecell}
\usepackage{slashed}
\usepackage[utf8]{inputenc}
\usepackage{xcolor}
\usepackage{subfigure}

\begin{document}

\title{New searches at reactor experiments based on the dark axion portal}

\author{Patrick deNiverville}
\affiliation{T2, LANL, Los Alamos, New Mexico 87545, USA}
\author{Hye-Sung Lee}
\affiliation{Department of Physics, KAIST, Daejeon 34141, Korea}
\author{Young-Min Lee}
\affiliation{Department of Physics, KAIST, Daejeon 34141, Korea}

\date{\today}

\begin{abstract}
A nuclear reactor is a powerful tool to study neutrinos and light dark sector particles.
Some reactor experiments have already proven to be extremely useful in these searches.
Considering the great interest in the power of the Intensity Frontier to search for new light particles, it would be desirable to explore the possibility of exploiting the existing reactor power sources for particle physics research.
We suggest a new reactor experiment searching for the dark sector.
The dark photon can be produced in a reactor core and decay into a photon and an axion in the presence of the dark axion portal through an axion-photon-dark photon vertex.
We investigate the potential to search for this new vertex with a monophoton signature and present the expected sensitivities at some of the existing reactor neutrino experiment detectors.
\end{abstract}

\maketitle

\section{Introduction}
Today, there are more than 400 operational nuclear power reactors in the world \cite{pris}. Tremendously intense physical reactions occur inside each reactor's core as it exploits the chain reaction of nuclear fission to generate power.
For instance, a typical 1~GW power reactor emits about $10^{20}$ antineutrinos from beta decay every second.

These reactors can be excellent sources of new light particles.
Motivated by particle dark matter, there has been a significant increase in the study of the various kinds of hypothetical new light particles such as the axion, light dark matter, the dark photon, and light sterile neutrinos \cite{Essig:2013lka}.
Studying these light dark sector particles does not require large energies but instead enormous intensities.
Thus nuclear reactors can be an ideal place to produce them, and many experiments have utilized the large intensity from reactors.

Ever since the first experimental confirmation of the neutrino's existence using the reactors at the Savannah River Plant~\cite{Cowan:1992xc}, reactors played a vital role in neutrino studies as a low energy and high intensity neutrino source.
Early experiments in the 1970s and the 1980s contributed to the understanding of the reactor neutrino flux, then Chooz and Palo Verde in the 1990s contributed to the knowledge of detector systematics and backgrounds~\cite{Cao:2017drk}.
KamLAND combined their data with the solar neutrino data, giving the best fit of $\Delta m_{21}^2=7.59 \times 10^{-5}$ eV$^2$ and $\tan^2\theta_{12}=0.47$~\cite{Abe:2008aa}.
Daya Bay~\cite{An:2012eh}, Double Chooz~\cite{Abe:2011fz} and RENO~\cite{Ahn:2012nd} measured $\theta_{13}$, and recently the most precise result provides $\sin^22\theta_{13}=0.0841$ with $|\Delta m_{ee}^2|=2.50\times10^{-3}$ eV$^2$~\cite{An:2016ses}.
Now, reactor neutrino oscillation experiments have entered into the precision era~\cite{Cao:2017drk}.

New physics searches in reactor neutrino oscillations, such as the sterile neutrino effect, are to be performed with the present and future reactor experiments such as TEXONO~\cite{Wong:2016lmb}, NEOS~\cite{Kim:2015qlu,Ko:2016owz,Siyeon:2017tsg}, and JUNO~\cite{An:2015jdp}.
Another branch of the reactor neutrino experiments is the coherent elastic neutrino-nucleus scattering (CE$\nu$NS) experiment.
It was first measured by the COHERENT collaboration using the accelerator at Oak Ridge National Laboratory in 2017~\cite{Akimov1123}.
Now, a series of reactor experiments for CE$\nu$NS are on in progress, including MINER~\cite{Agnolet:2016zir}, CONUS~\cite{Buck_2020}, CONNIE~\cite{Aguilar_Arevalo_2016}, and $\nu$-cleus~\cite{Strauss_2017}.

The early reactor axion experiments contributed to the exclusion of the original QCD axion model mostly through the decay of axions~\cite{Vuilleumier:1981dq,DATAR198263,Ketov:1986az,Koch:1986aq,Altmann:1995bw}.
The global Peccei-Quinn (PQ) symmetry was proposed to solve the strong \textit{CP} problem~\cite{Peccei:1977hh,Peccei:1977ur}, predicting the existence of a pseudo-Goldstone boson, the axion ~\cite{Weinberg:1977ma,Wilczek:1977pj}.
The first realization of the axion was the Peccei-Quinn-Weinberg-Wilczek (PQWW) axion with its symmetry breaking at the electroweak scale.
The signal of an axion decaying into two photons was investigated at the Institut Laue-Langevin (ILL) reactor~\cite{Vuilleumier:1981dq} and at the nuclear power reactor Biblis A~\cite{Koch:1986aq}.
The $a\rightarrow e^+e^-$ decay was searched for at the Bugey nuclear power reactor 5~\cite{Altmann:1995bw} while the diphoton signal from axion produced by neutron capture $n+p \rightarrow d+a$ was looked for at a 500 MW light-water power reactor at Tarapur atomic power station~\cite{DATAR198263}.

The invisible axion models were introduced with a symmetry breaking scale larger than the electroweak scale to avoid existing experimental constraints.
One type is known as the Kim-Shifman-Vainshtein-Zakharov (KSVZ)~\cite{Kim:1979if,SHIFMAN1980493} and another as Dine-Fischler-Srednicki-Zhitnisky (DFSZ)~\cite{DINE1981199}.
They were also investigated using neutron capture and nuclear transition as axion production mechanisms~\cite{Chang:2006ug}, giving the constraints on the couplings $G_{a\gamma\gamma}$ and $G_{aee}$ which rule out DFSZ and KSVZ models for $10^4\,\mathrm{eV} \lesssim m_a \lesssim 10^6\,\mathrm{eV}$.
Recently, there are also studies to search for axion-like particles (ALPs) utilizing the reactor neutrino experiments~\cite{Dent:2019ueq,AristizabalSierra:2020rom}.
ALP-photon, ALP-electron and ALP-nucleon couplings were explored considering most of the channels including Primakoff and Compton-like processes, nuclear de-excitation and axio-electric absorption as well as decay processes~\cite{AristizabalSierra:2020rom}.

Dark photon production in a nuclear reactor and its subsequent detection through either scattering or decay to dark sector particles was previously studied in Refs.~\cite{Park:2017prx,Danilov:2018bks,Ge:2017mcq}.
The dark photon has motivations from both dark matter related (such as the dark matter annihilation into dark photons to explain the positron excess~\cite{ArkaniHamed:2008qn}) and unrelated (such as the muon $g-2$ anomaly~\cite{Gninenko:2001hx,Fayet:2007ua,Pospelov:2008zw}) phenomena.
Dark photons produced and detected by (inverse) Compton-like processes in reactor neutrino experiments give constraints on the kinetic mixing parameter of $\varepsilon<2.1\times 10^{-5}$ for TEXONO and $\varepsilon<1.3\times 10^{-5}$ for NEOS with 95\% C.L. for a sub-MeV dark photon~\cite{Park:2017prx}.
Below the resonance point of $m_{\gamma'}\simeq20$\,eV the sensitivity to the dark photon decreases as $m_{\gamma'}$ decreases~\cite{Danilov:2018bks}.

In this paper, a new search at reactor experiments exploiting a monophoton as a decay product of the dark photon generated in the reactor core is suggested as a search for the dark axion portal.
We take the RENO, NEOS, MINER, and CONUS experiments as our example setups for the numerical studies among many existing and planned reactor experiments.
In Sec.~\ref{sec:portals}, we discuss the relevant portals including the dark axion portal and the vector portal. We then describe the method we use to evaluate the number of signal events in Sec.~\ref{sec:DAPtest}, and provide an analytic expression one can use to make a rough estimation of the number of signal events for a given experiment in Sec.~\ref{sec:benchmark}.
In Sec.~\ref{sec:signals}, we show the results of the feasibility study for the example setups, present an optimal design, compare it with fixed target neutrino experiments and discuss astrophysical constraints.
We study the effect of an additional vector portal on top of the dark axion portal in Sec.~\ref{sec:VP}, including the implications of additional light dark particles.
Finally, we summarize and discuss our findings by showing the summary plot in Sec.~\ref{sec:summary}.

\section{Relevant portals}
\label{sec:portals}

A ``portal" is a concept to connect the visible (standard model) sector and the dark sector, which helped to establish strategies for searching for the dark sector particles.
Several portals can connect to the photon including the axion portal, the dark axion portal and the vector portal.

The axion can couple to standard model particles via the axion portal given by
\begin{equation}
    \mathcal{L}_\text{axion portal} = \frac{G_{agg}}{4}aG_{\mu\nu}\tilde G^{\mu\nu} + \frac{G_{a\gamma\gamma}}{4}aF_{\mu\nu}\tilde F^{\mu\nu} + \cdots,
\end{equation}
where $F_{\mu\nu}$ and $G_{\mu\nu}$ are the field strength of the photon and the gluon, respectively, the tilde denotes the dual field strength and $a$ is the axion field.

If the axion and dark photon coexist, they can also couple together, giving the dark axion portal as~\cite{Kaneta:2016wvf}
\begin{equation}
    \mathcal{L}_\text{dark axion portal} = \frac{G_{a\gamma'\gamma'}}{4}aZ'_{\mu\nu}\tilde Z'^{\mu\nu} + \frac{G_{a\gamma\gamma'}}{2}aF_{\mu\nu}\tilde Z'^{\mu\nu},
\end{equation}
where $Z'^{\mu\nu}$ is the field strength of the dark photon.
Though the first term, an axion coupling to two dark photons, is not the usual portal relating the visible and dark sectors, once the second term, axion-photon-dark photon coupling, is introduced, the first term is inevitable.
Note that the second term is not simply a combination of the vector and axion portal, but rather exploits the dark gauge couplings \cite{Kaneta:2016wvf}.
See Refs.~\cite{Choi:2016kke,Kaneta:2017wfh,Daido:2018dmu} for more about the dark axion portal.

The axion in the axion and the dark axion portals could be the QCD axion, which explains the strong \textit{CP} problem, as well as a more general axion-like particle (ALP), which does not necessarily address the strong \textit{CP} problem.
In the parameter range we take for the analysis, it should more properly be called the ALP, but we refer to it simply as an axion throughout this paper.

The vector portal represents the mixing between two $U(1)$ gauge symmetries~\cite{Holdom:1985ag}, which is given by
\begin{equation}
    \mathcal{L}_\text{vector portal} = \frac{\varepsilon}{2}F_{\mu\nu}Z'^{\mu\nu}
\end{equation}
after the electroweak symmetry breaking. The kinetic mixing parameter $\varepsilon$, which controls the amount of the mixing, needs to be very small to evade experimental constraints~\cite{Essig:2013lka}.

\begin{figure*}[tb]
\centering
\subfigure[]{\includegraphics[width=0.4\textwidth]{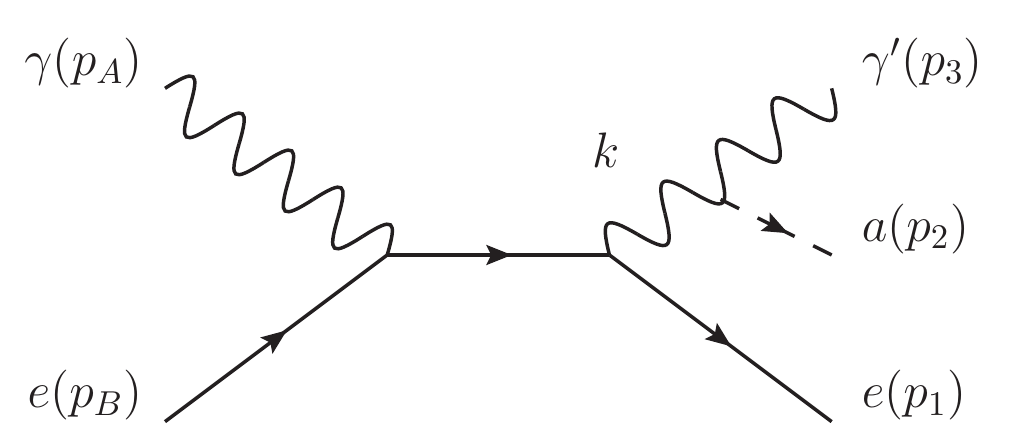}} ~~~
\subfigure[]{\includegraphics[width=0.3\textwidth]{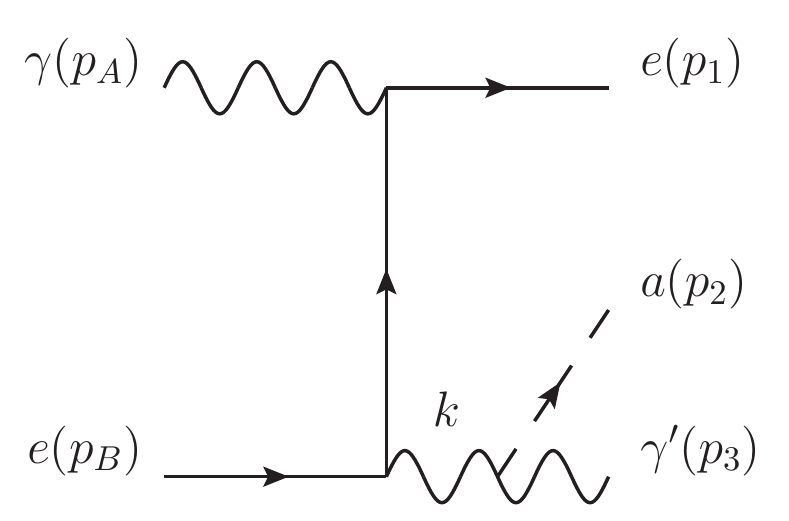}}
\caption{Feynman diagrams for the process $\gamma e \to \gamma^\prime a e$ using the dark axion portal $G_{a\gamma\gamma'}$.
}
\label{fig:feynman_DAP}
\end{figure*}

\section{Testing the dark axion portal at reactors}
\label{sec:DAPtest}

\subsection{Dark photon production spectrum from the reactor core}

The reactor photon production distribution was modeled by
\begin{equation}
 \label{eq:reactor_photon}
 \frac{dN_\gamma}{dE_\gamma} = \frac{0.58 \times 10^{18}}{\mathrm{sec\cdot MeV}} \, \frac{P}{\mathrm{MW}} \,  e^{- E_\gamma/ (0.91 \mathrm{MeV})},
\end{equation} 
where $P$ is the thermal power of the reactor and $E_\gamma$ is the photon energy~\cite{Bechteler:1983}.
Dark photons can be produced in nuclear reactors through a number of processes, but we focus on $\gamma e \to \gamma^\prime a e$ through the dark axion portal (see Fig.~\ref{fig:feynman_DAP}).
We will only study the production and detection of the dark photon through its decay in reactor neutrino experiments in detail in this work.

The axion could potentially be detected through scattering or, if sufficiently massive, its decay.
Recent analysis of reactor neutrino experiments shows that they could place limits that beat $G_{a\gamma\gamma} = 10^{-6}\,\textrm{GeV}^{-1}$ for some masses~\cite{Dent:2019ueq}.
The axion produced through the decay of the dark photon could also be detected, but the mean free path of the axion in the detector material is much larger than the size of the detector, e.g., $\mathcal{O}(10)$\,m for Germanium and $\mathcal{O}(100)$\,m for typical liquid scintillator solvent even in the case of $G_{a\gamma\gamma}=1\,\textrm{GeV}^{-1}$~\cite{AristizabalSierra:2020rom,Ahn:2010vy}.
We also consider the axion much lighter than the dark photon, which leads to an effectively large decay length, e.g., $\mathcal{O}(100)$\,km for $m_a=0.1\,\textrm{keV}$ with $E_a=1\,\textrm{MeV}$ and $G_{a\gamma\gamma}=1\,\textrm{GeV}^{-1}$~\cite{Dent:2019ueq}.
Therefore, the signal from the axion would clearly be subdominant.

The vector portal production through kinetic mixing has been studied previously in Refs. \cite{Park:2017prx, Ge:2017mcq, Danilov:2018bks}, and we will adopt a similar approach for production through the dark axion portal.
Assuming that Compton scattering is the dominant process, the number and spectrum of $\gamma^\prime$ produced by reactor photons can be calculated as 
\begin{equation}
 \label{eq:dp_rate_dap}
 \frac{dN_{\gamma^\prime}}{dE_{\gamma^\prime}} = \int dE_\gamma \frac{1}{\sigma_{\mathrm{tot}}} \frac{d\sigma_{\gamma e\to a e\gamma^\prime}}{dE_{\gamma'}} \frac{dN_\gamma}{dE_\gamma},
\end{equation}
where $\sigma_\mathrm{tot}$ is the total interaction cross section between photons and matter and we integrate the energy over the range $[0,15]$ MeV. 

\begin{figure}[b]
 \centerline{\includegraphics[trim={0cm 0cm 0cm 0cm},clip,width=0.48\textwidth]{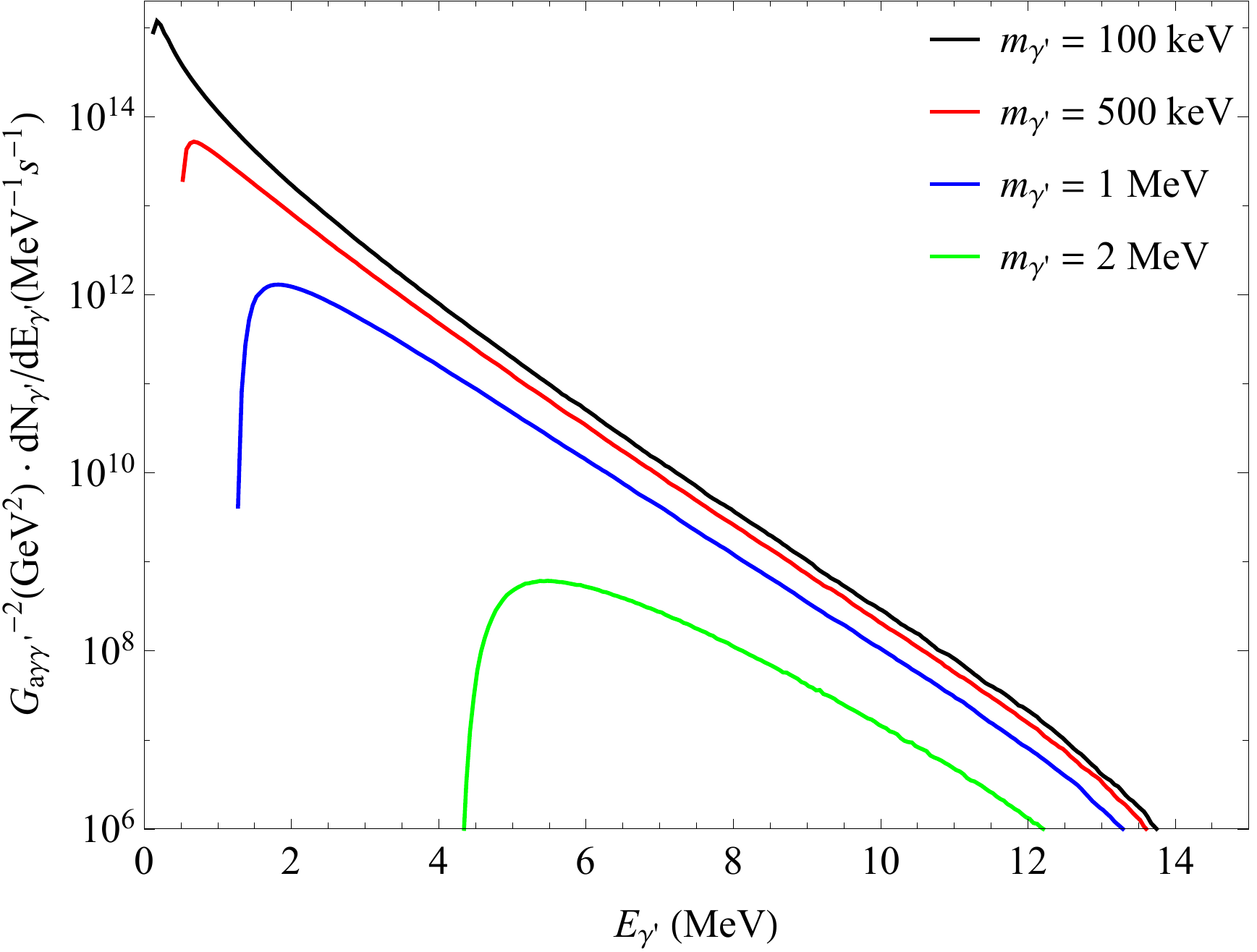}
}
 \caption{Production spectrum of dark photons through the dark axion portal ($G_{a\gamma\gamma'}$) in a 1 GW nuclear reactor for four dark photon masses. One can see that there are kinematic cutoffs in the energy spectrum that depend on the dark photon mass.}
 \label{fig:production_dap}
\end{figure}

The cross section of $\gamma e \to \gamma^\prime a e$ can be found with the following amplitude, evaluated with the assistance of FeynCalc \cite{Mertig:1990an,Shtabovenko:2016sxi,Shtabovenko:2020gxv}:
\begin{eqnarray}
\label{eq:reactor_prod}
 \mathcal{M} &=& \mathcal{M}_s + \mathcal{M}_u \nonumber \\
 \mathcal{M}_s &=&  \frac{e^2 G_{a\gamma \gamma^\prime}}{k^2 (s-m_e^2)} k^\sigma p_3^\lambda \epsilon_{\sigma\mu \lambda \delta} \epsilon^\nu(p_A) {\epsilon^*}^\delta(p_3) \nonumber \\
 &\times& \bar u (p_1,m_e) \gamma^\mu 
 ((\slashed p_A+\slashed p_B)+m_e) \gamma^\nu u(p_B,m_e), \\
 \mathcal{M}_u &=&  \frac{e^2 G_{a\gamma \gamma^\prime}}{k^2 ((p_B-k)^2-m_e^2)} k^\sigma p_3^\lambda \epsilon_{\sigma \mu \lambda \delta} \epsilon^\nu(p_A) {\epsilon^*}^\delta(p_3) \nonumber \\
 &\times& \bar u (p_1,m_e) \gamma^\nu 
 ((\slashed p_B-k)+m_e) \gamma^\mu u(p_B,m_e).
\end{eqnarray}
The $\gamma^\prime$ emission spectrum for production through the dark axion portal is shown in Fig. \ref{fig:production_dap}. 

\subsection{Dark photon decay events expected in the detector}

So long as $m_{\gamma^\prime}< 2 m_e$, the dominant decay is through $\gamma^\prime \to a \gamma$, resulting in a monophoton signal potentially detectable in neutrino detectors. The width of the dark photon is given by \cite{Kaneta:2017wfh}
\begin{equation}
 \Gamma_{\gamma^\prime \to a \gamma} = \frac{1}{96 \pi} G_{a\gamma\gamma^\prime}^2 m_{\gamma^\prime}^3 \left(1 - \frac{m_a^2}{m_{\gamma^\prime}^2}\right)^3.
\end{equation}
In our analysis we take $m_a$ much less than $m_{\gamma'}$, which effectively leads to $m_a/m_{\gamma'} \simeq 0$.

The event rate in a distant detector is calculated by two different methods:
\begin{enumerate}
 \item Integration over the production distribution convolved with the decay probability for a distant detector.
 \item A Monte Carlo of the production, propagation and decay of a dark photon with a detector.
\end{enumerate}
The integration approach requires a number of assumptions and approximations, but so long as they are satisfied it agrees well with the results of the Monte Carlo approach.

Let us consider a detector of volume $V$ located at some distance $L$ from the source.
For the numerical approximation, we will assume that the production is isotropic, and the decay rate of incident dark photons in the detector is constant over the entire volume of the detector, which requires that $\sqrt[3]{V} \ll L$ and that the mean travel distance before decay $\beta \tau_{\gamma^\prime} c \gtrapprox \sqrt[3]{V}$.
With these assumptions satisfied, the exact geometry and location of the detector can be ignored, as they will not affect the observed event rate.
We can instead model the detector volume as a spherical shell of radius $L$ and thickness $2 \times \delta L$ where
\begin{equation}
 \delta L = - \frac{4 L^2}{\ell} + \ell
\end{equation}
with
\begin{equation}
    \ell = \left(\frac{3 V + \sqrt{256 L^6 \pi^2+9 V^2}}{2\pi} \right)^{1/3}.
\end{equation}
The event rate in a detector located at some distance $L$ can be calculated numerically as
\begin{eqnarray}
 N_\mathrm{decay} &=& \textrm{Br}(\gamma'\to a \gamma) \, T \, \int dE_{\gamma^\prime} \, \frac{dN_{\gamma^\prime}}{dE_{\gamma^\prime}} \, f_\mathrm{min} \nonumber \\
 &\times& \left( \exp\left(-\frac{L-\delta L}{c\beta\gamma\tau}\right) - \exp\left(-\frac{L+\delta L}{c\beta\gamma\tau}\right)\right),
 \label{eq:event_rate}
\end{eqnarray}
where $\beta \gamma = \frac{p_{\gamma^\prime}}{m_{\gamma^\prime}} = \frac{\sqrt{E_{\gamma^\prime}^2-m_{\gamma^\prime}^2}}{m_{\gamma^\prime}}$, $\tau = \hbar \Gamma^{-1}$ is the lifetime of the dark photon, $T$ is the run time of the experiment, $\textrm{Br}(\gamma'\to a \gamma)$ is the branching ratio and $f_\mathrm{min}$ is the fraction of decays $\gamma^\prime \to a \gamma$ that produce photons with an energy above some cut $E_\mathrm{min}$.
We calculate this fraction by determining the $\cos \theta$ of a decay product of the dark photon with energy $E_\mathrm{CM}$ in the dark photon rest frame:
\begin{equation}
 \cos \theta = \frac{1}{\beta p_\mathrm{CM}}\left( \frac{E_\mathrm{min}}{\gamma} - E_{CM} \right),
\end{equation}
where
\begin{equation}
p_\mathrm{CM}=\frac{m_{\gamma'}^2-m_a^2}{2 m_{\gamma^\prime}} \simeq \frac{m_{\gamma'}}{2}
\end{equation}
is the momentum of the decay products in the rest frame of the dark photon, $\beta= p_{\gamma^\prime,\mathrm{lab}}/E_{\gamma^\prime,\mathrm{lab}}$ is the  boost required to switch from the rest frame of the dark photon to the lab frame.
The fraction of photons with energies above $E_\mathrm{min}$ is therefore
\begin{equation}
 f_\mathrm{min} = \frac{1-\cos \theta}{2}.
\end{equation}

We also generated results using a modified version of the \texttt{BDNMC} code \cite{deNiverville:2016rqh}\footnote{Our Monte Carlo code is available at \url{https://github.com/pgdeniverville/BdNMC}.}. 
Details of the implementation specific to the dark axion portal can be found in Refs \cite{deNiverville:2018hrc,deNiverville:2019xsx}.
Sample files of dark photons were generated from the $\frac{dN_{\gamma^\prime}}{dE_{\gamma^\prime}}$ distribution shown in Eq.~\eqref{eq:dp_rate_dap}.
The results of the Monte Carlo approach were in good agreement with our semianalytical method.

\begin{figure*}[t]
\centering
\subfigure[]{\includegraphics[width=0.48\textwidth]{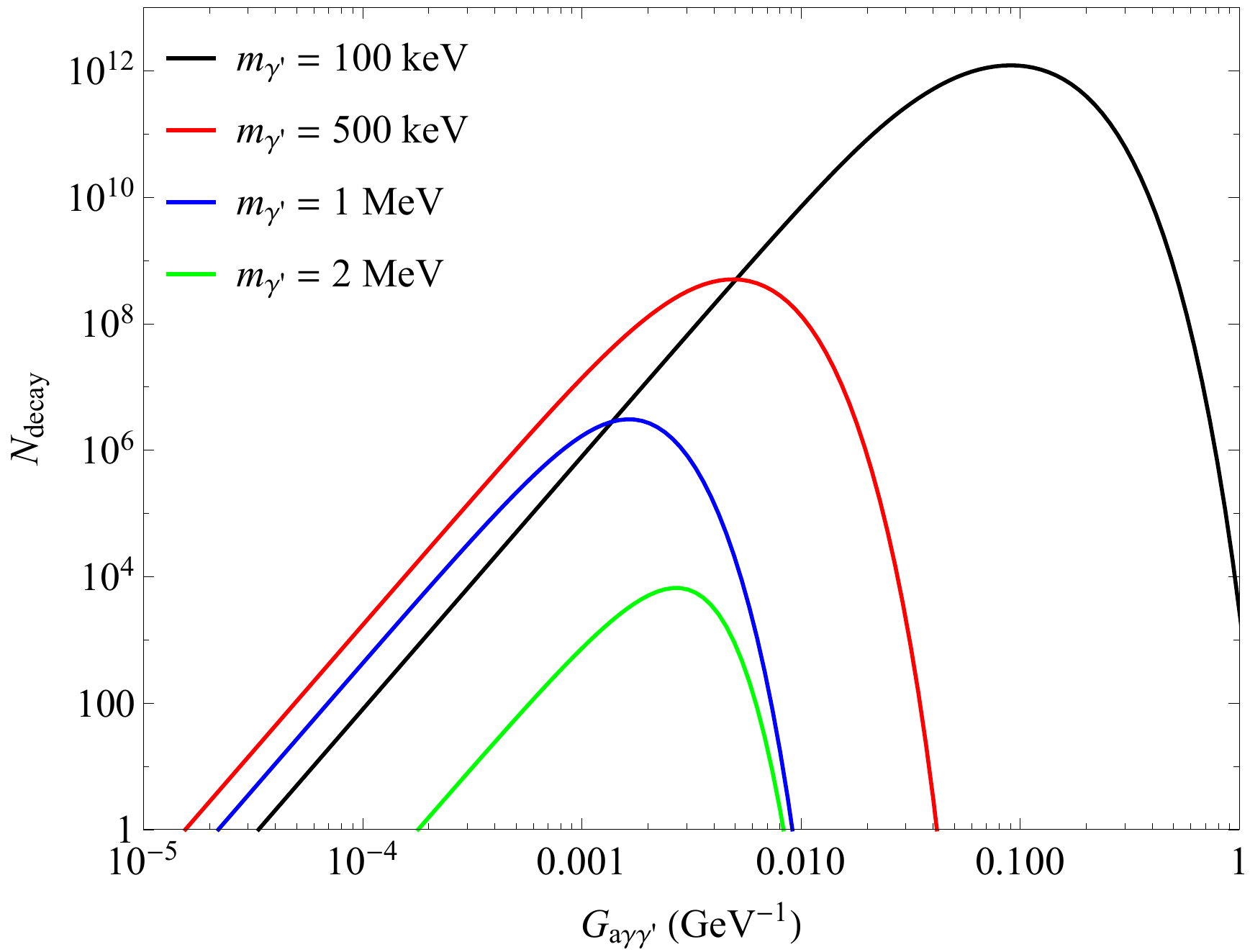}
\label{fig:benchmark_a}}
~~~
\subfigure[]{\includegraphics[width=0.48\textwidth]{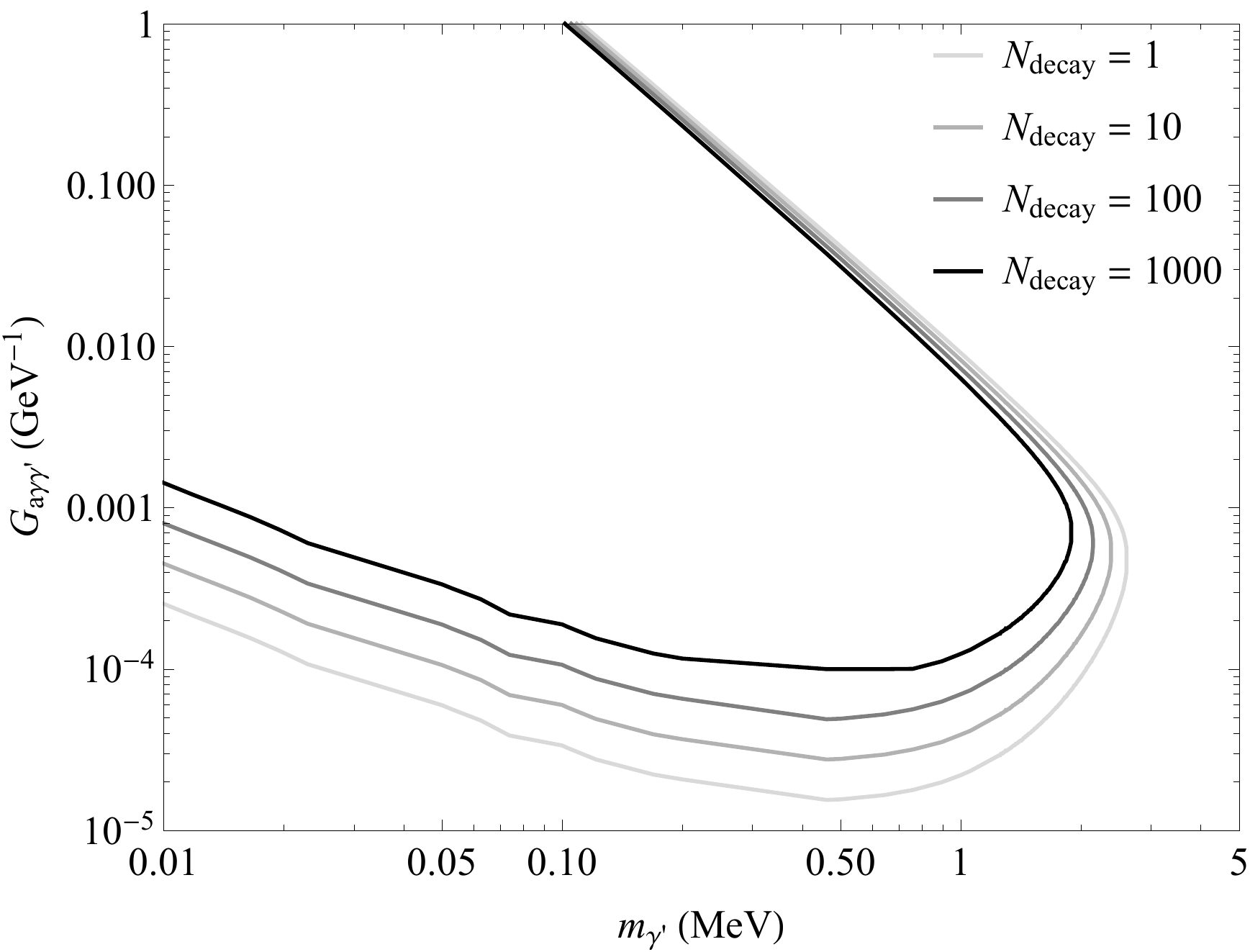}
\label{fig:benchmark_b}}
\caption{The expected number of dark photon decay events under the benchmark setup with one detector of 1\,m$^3$ 100\,m far from the 1\,GW single reactor core with 1\,year of run time.
There are more than enough events to be detected due to the high intensity of the flux from the reactor.
The dark photon production becomes minute to be detected if the dark axion portal coupling is too small, while most of them decay before they reach the detector if the coupling is too large.
(a) The number of decays as a function of the coupling for various dark photon masses.
(b) The number of decays as a contour in the coupling-dark photon mass space.
}
\label{fig:benchmark}
\end{figure*}

\section{Benchmark chart for dark photon events}
\label{sec:benchmark}
Every reactor experiment has a different number of reactors with their own thermal power and detector configurations with disparate detection techniques and backgrounds.
We studied a benchmark setup and present the results so that one can roughly convert our results to the expected signals for different setups.

Figure~\ref{fig:benchmark} shows the expected number of dark photon events under the benchmark setup of 1\,GW single reactor core with a 1\,m$^3$ detector 100\,m from the reactor with 1\,year of run time.
In Fig.~\ref{fig:benchmark_a}, the events show a peak as the dark axion portal coupling increases.
This is because dark photon production becomes too weak if the coupling is very small, while they mostly decay too early before they reach the detector if the coupling is too large.
This is also shown in Fig.~\ref{fig:benchmark_b} as upper and lower parts of the contours.

As a quick estimation for the event numbers of one's own experimental setup, we present a simple conversion equation from the benchmark setup as
\begin{eqnarray}
\label{eq:estimation}
&N_\text{decay}&(P, V, T, L) \nonumber \\
&=& N^\text{benchmark}_\text{decay} \left( \frac{P}{1\text{\,GW}} \right) \left( \frac{V}{1 \text{\,m}^3} \right) \left( \frac{T}{1\text{\,year}} \right) \nonumber\\
&\times& \left( \frac{100\text{\,m}}{L} \right)^2 \exp\left(-\frac{L-100\text{\,m}}{c\tau}\right) ~,
\end{eqnarray}
where $N^\text{benchmark}_\text{decay}$ is $N_\textrm{decay}$ in Fig.~\ref{fig:benchmark}, $P$ is the power of the reactor, $V$ is the volume of the detector, $T$ is the runtime of the experiment, $L$ is the distance between the reactor core and the detector and $\tau$ is the lifetime of the dark photon.
The dark photon is assumed to travel at approximately the speed of light, and note that the lifetime depends on the dark photon mass and the coupling.
Also, the volume is assumed to be linearly proportional, which can be quite correct as long as the detector has small volume compared to the distance.
One can use Eq.~\eqref{eq:estimation} to give a rough estimation of the sensitivity for a given experimental setup.

Under isotropic production, the flux decreases with travel length as $1/L^2$ if the particles do not decay at an appreciable rate during their flight.
When there is a decay channel, the flux will decrease by a factor of $\exp(-\frac{L}{c \beta \gamma \tau})/L^2$, which is also depicted in Eq.~\eqref{eq:estimation}.
A comparison of the fluxes at multiple distances  provides a method of detecting a decay process.

\begin{table*}[bt]
\caption{Summary of the experimental setups. The specifications for the experiments are based on Refs.~\cite{Ko:2016owz,Ahn:2010vy,AristizabalSierra:2019hcm,Agnolet:2016zir,Buck_2020}, and the background rates are determined based on Refs.~\cite{Ahn:2010vy,Dent:2019ueq,AristizabalSierra:2019hcm,Yoomin:2018}.
The detector volume of CONUS and MINER was estimated from their payload.
}
\renewcommand{\tabcolsep}{2.6mm}
\centering
\begin{tabular}{c|c|c|c|c|c}
\Xhline{3\arrayrulewidth}
Experiment & \makecell{Detector\\volume} & \makecell{Reactor\\power} & \makecell{Reactor-detector\\distance} & \makecell{Background\\rate} & Energy cutoff \\ \Xhline{2\arrayrulewidth}
	CONUS	&	751.46\,cm$^3$	&   3.9\,GW & 17\,m	& 12\,Hz & Negligible \\ \hline
	MINER\footnote{Phase-2 is assumed.}	&	3085.2\,cm$^3$	& 1\,MW	   & 2.835\,m	& 6\,Hz & Negligible \\ \hline
	RENO	&	18.7\,m$^3$	& \makecell{16.4\,GW (total)\\2.73\,GW (each)}	   & \makecell{304.8\,m (nearest)\\739.1\,m (farthest)}	& 30\,Hz & 1\,MeV\\ \hline
	NEOS	&	1.008\,m$^3$	& 2.73\,GW   & 23.7\,m	& 0.16\,Hz\footnote{Signal+Background rate.} & 3.5\,MeV\\ \Xhline{3\arrayrulewidth}
\end{tabular}
\label{tab:setup}
\end{table*}

\section{Signals in reactor neutrino experiments}
\label{sec:signals}

\subsection{Experimental setup}

As example experiments, we consider RENO (near detector only), NEOS, MINER, and CONUS among many other reactor experiments.
RENO (Reactor Experiment for Neutrino Oscillation) is an experiment to measure the neutrino oscillation parameter $\theta_{13}$ and NEOS (NEutrino Oscillation at Short baseline) searches for the sterile neutrino.
MINER (Mitchell Institute Neutrino Experiment at Reactor) and CONUS (COherent elastic NeUtrino nucleus Scattering) are to measure the coherent elastic neutrino-nucleus scattering.
The information related to the calculation of event rates is summarized in Table~\ref{tab:setup}.

Both RENO and NEOS are located at the Hanbit Nuclear Power Plant in Yeonggwang, Republic of Korea, and use liquid scintillator detectors.
The near detector of RENO has a volume of 18.7 $\mathrm{m}^3$ and is located near six nuclear reactors with a combined power output of 16.4 GW~\cite{Ahn:2010vy}.
The reactor-detector distances are 304.8 m, 336.1 m, 451.8 m, 513.9 m, 667.9 m, and 739.1 m.
The NEOS detector has a volume of 1.008\,$\mathrm{m}^3$ and is located 23.7 meters from the fifth reactor~\cite{Ko:2016owz}.
The total observed decay signal can be found by the simple addition of the individual signals from each reactor, though only the nearest reactors contribute significantly.

MINER is located at the Nuclear Science Center at Texas A\&M University utilizing a 1\,MW TRIGA (Training, Research, Isotopes, General Atomics) nuclear reactor.
Phase-1 of MINER is a demonstration experiment, hence we consider phase-2, with 10 times less background and 10 times larger payload~\cite{AristizabalSierra:2019hcm}.
The detector consists of cryogenic germanium and silicon detectors expected to have a threshold of around 100\,eV.
It is 2.835\,m far from the reactor core and the volume is 3085.2\,cm$^3$ with a 20\,kg payload of Ge/Si~\cite{Agnolet:2016zir,AristizabalSierra:2019hcm}.
Here, we approximated the detector volume from the mass of the payload, and the core proximity is calculated based on the setup in Ref.~\cite{Agnolet:2016zir}.

CONUS is located at the commercial nuclear power plant of Brokdorf, Germany with 3.9\,GW thermal power~\cite{Buck_2020}.
It has four germanium detectors expected to have a threshold around $300$\,eV.
The detector is 17\,m far from the reactor core and the volume is approximately 751.46\,cm$^3$ with a 4\,kg payload of Ge.
The detector volume is calculated in the same way as MINER.

There are various single photon background sources from the radioactive isotopes in the nearby rocks, PMT (photomultiplier tube) glass, liquid scintillator and so on~\cite{Ahn:2010vy}.
The background rate for RENO is calculated from the measurement of the isotope concentration and the simulation of detector acceptance in Ref.~\cite{Ahn:2010vy}, giving a single photon rate of 30\,Hz with energy above 1\,MeV.
RENO requires a cutoff of 1\,MeV since it looks for inverse beta decay, and the prompt signal of the positron has minimum energy of 1.022\,MeV.

The single event rate in the NEOS detector caused from alpha and beta particles, neutrons, and gammas is measured and reported in Ref.~\cite{Yoomin:2018}, and it is hard to separate single gamma events from the other single event backgrounds.
Therefore, we conservatively take the single event rate to be the total rate including both signals and backgrounds.
The energy cutoff is applied at 3.5\,MeV as the measurement is unreliable below 3.5\,MeV, and it also removes the huge backgrounds from radioactive isotopes at low energies.
The single event rate during reactor-on period were measured to be 0.16\,Hz.
Discriminating between gammas and other particles would further reduce the backgrounds, increasing the significance of the result.

The background rates for MINER and CONUS are adopted from Ref.~\cite{Dent:2019ueq}, assuming a uniform spectrum up to 2.6\,MeV, where the radiation from the radioactive isotopes rapidly diminishes (see Fig.~11 of Ref.~\cite{Apollonio:2002gd}).
The background rate is 100\,kg$^{-1}$keV$^{-1}$day$^{-1}$ for CONUS and 10\,kg$^{-1}$keV$^{-1}$day$^{-1}$ for the phase-2 of MINER.
These correspond to 12\,Hz for CONUS and 6\,Hz for MINER considering their payload and the energy range of interest.

The significance of standard deviations, given as
\begin{equation}
 \frac{N_s}{\sqrt{N_s+N_b}} \textrm{ ($N_s$: signal, $N_b$: background),}
\end{equation}
is calculated to obtain $N_s$ giving $2\,\sigma$ ($95\%$ C.L.) contour.
Since the radiation from the isotopes peaks at some specific energies~\cite{Apollonio:2002gd}, detailed background analysis by the collaboration to reduce those peaks might enhance the sensitivities significantly.

\begin{figure}[b]
\centering
\includegraphics[width=0.48\textwidth]{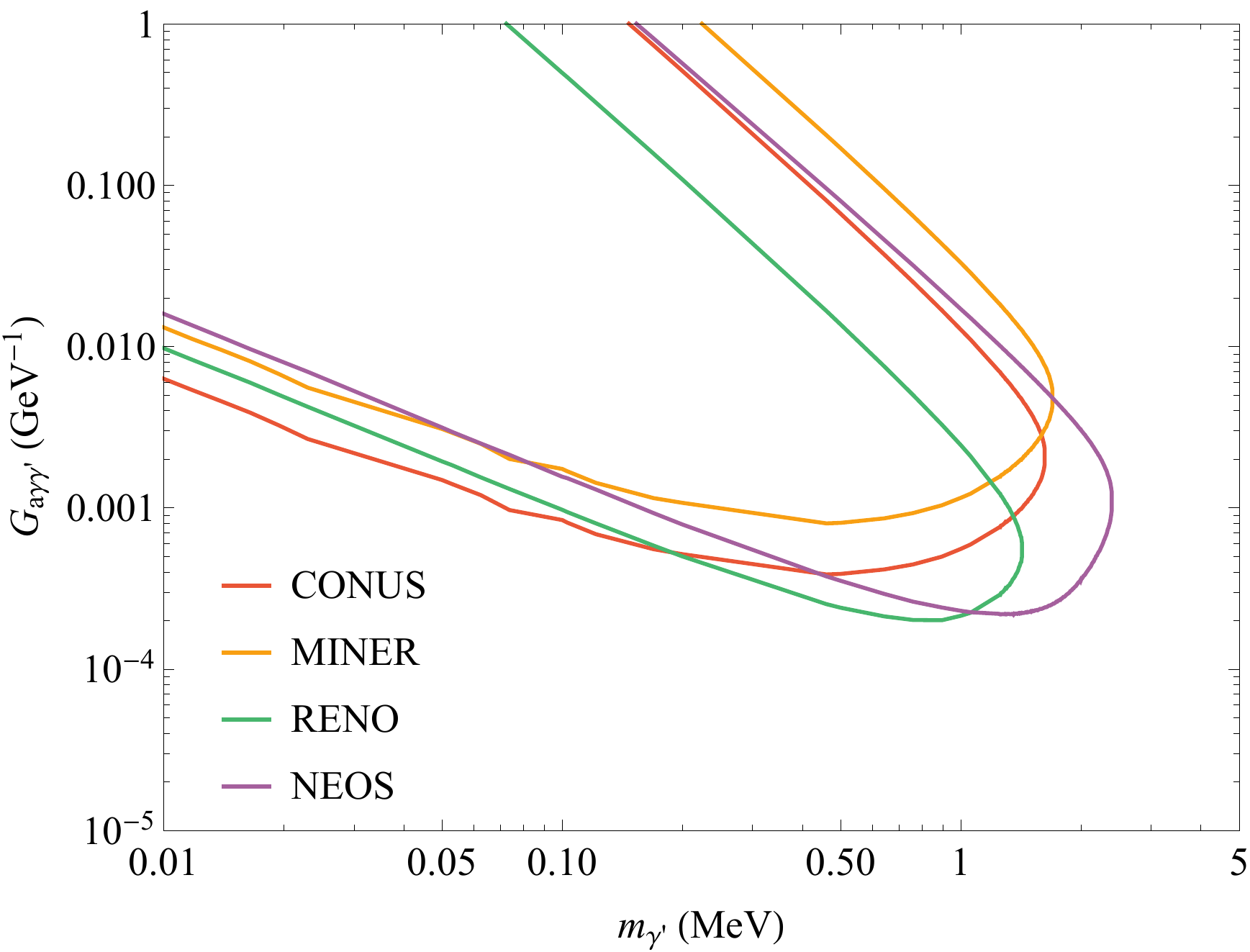}
\caption{Expected sensitivities at CONUS, MINER, RENO and NEOS.
Presented are 95\%\,C.L. contours for one year of data.
The result with 1\,MeV-cutoff is shown for RENO, and the result with 3.5\,MeV-cutoff is shown for NEOS.
MINER and CONUS do not require an energy cutoff.
The cutoff reduces the sensitivity at the lower bound of $G_{a\gamma\gamma'}$ since the dark photon signal mostly comes from the low energy region when the coupling is small (see Fig.~\ref{fig:production_dap}).
NEOS has better sensitivity than RENO in larger couplings and masses benefiting from its close location to the reactor core.
MINER and CONUS have smaller coverage compared to RENO and NEOS because of their smaller detector volume.
An analysis of the background energy spectrum could improve the presented sensitivities.}
\label{fig:dap}
\end{figure}

\subsection{Expected sensitivities}
Presented in Fig.~\ref{fig:dap} are the expected sensitivities at CONUS, MINER, RENO, and NEOS in the dark photon mass ($m_{\gamma'}$) and dark axion portal coupling ($G_{a\gamma\gamma'}$) parameter space.
Contours of 95\% C.L. (2$\sigma$) are shown for each experiment with one year of data.
Due to its nearer location to the reactor, NEOS is capable of probing shorter dark photon lifetimes, and therefore larger couplings and masses compared to RENO.
MINER and CONUS have smaller coverage because of their smaller detector size.

A 1\,MeV-cutoff is applied to RENO in order to control backgrounds, and NEOS has 3.5\,MeV-cutoff as mentioned before.
MINER and CONUS need no cutoff since their detectors are sensitive enough to detect all photons resulting from the decay of reactor dark photons.
When there is a cutoff, the coverage in the small coupling region (the lower part of the contour) is reduced since most of the contribution in the dark photon flux comes from the low energy region which is thrown away (see Fig.~\ref{fig:production_dap}), and it is especially critical for small couplings.

Some comments about the optimal experiment design are in order.
To explore the lowest coupling region, as Eq.~\eqref{eq:estimation} implies, a larger volume is best.
Yet, the dependence on the distance to the detector can also play a vital role because of the isotropic production of the flux.
Thus an obvious improvement could be achieved if we have a large detector, say RENO size, close to the reactor core, say the NEOS distance.
\begin{eqnarray}
&N_\text{decay}(\text{RENO size, NEOS distance}) \nonumber \\
&\sim 18.6 N_\text{decay}(\text{NEOS})
\end{eqnarray}

One of the limiting factors of the sensitivity of a decay experiment is the volume in which the particle of interest may decay.
With relatively low cost, the effective volume, and therefore the sensitivity, of a decay experiment may be enhanced through the addition of an uninstrumented decay volume.
So long as the particle products then enter the instrumented volume of the detector, they can still be detected and contribute additional signals.
However, the reactor experiments often see little benefit from the addition of a decay volume as the energy of the reactor dark photon is a few MeV range.
The photons produced through dark photon decays have a large enough angular spread that very few are capable of reaching the active region of the experiment, instead of hitting the walls of the decay volume.
We numerically studied the effect of adding a cylindrical decay pipe to the RENO experiment, but a decay region of four times the detector volume did not show much greater sensitivity.
One can improve on the design by enlarging the radius of the decay pipe and the cross sectional area of active region of the detector. 

\subsection{Comparison with fixed target neutrino experiments}

Other neutrino experiments provide additional options for investigating the dark axion portal and are capable of probing different parameter space from reactor neutrino experiments.
A second type of neutrino experiment that we considered previously in Ref.~\cite{deNiverville:2018hrc} is the fixed target neutrino experiment (FTNE) which utilizes an accelerator to produce a high-intensity proton beam.
The protons impact thick targets and generate charged mesons such as pions and kaons, which then decay to neutrinos.
However, the neutral pseudoscalar mesons such as $\pi^0$ and $\eta$ are also produced in large numbers during the process.
The $\pi^0$ primarily decays to two photons, but the dark axion portal suggests an additional decay mode: $\pi^0\rightarrow \gamma+a+\gamma'$.
The $\eta$ can also produce a photon, a dark photon, and an axion in the same manner as the $\pi^0$.
Here, we will compare the ability to probe the dark axion portal between reactor experiments and FTNEs.
The dark photon and axion production rates would play a key role in comparing those two types of experiments.

Chosen as an example is the Deep Underground Neutrino Experiment (DUNE), a next-generation FTNE in development.
It plans to take advantage of $1.1\times10^{21}$ protons on target (POT) per year from the Long-Baseline Neutrino Facility in Fermilab~\cite{Abi:2020evt,Abi:2020qib}.
The production of $\pi^0$ and $\eta$ per POT are estimated to be 2.89 and 0.33 respectively, providing total $\sim10^{21}$ of $\pi^0$ and $\eta$ per year~\cite{Kelly:2020dda}.
Although the particles in the FTNEs are focused, the angular acceptance cannot be ignored; the angular acceptance is simulated to be about 0.5\% for DUNE and this should be taken into account.
On the other hand, the number of photons with energies in the range [0,\,15]\,MeV from a 1\,GW reactor is $\sim10^{28}$ per year.
The production is isotropic in the reactor experiments; in the case of $\mathcal{O}(1)\,\textrm{m}^2$ cross-sectional detector at $\mathcal{O}(10)\,\textrm{m}$ from the source, we would expect only $\frac{1}{4 \pi 10^2}$ of the particles produced to intersect with the detector.
The branching ratio of $\pi^0\rightarrow \gamma+a+\gamma'$ is suppressed by $e^4 G_{a\gamma\gamma'}^2$ which is the same for the ratio $\sigma_{\gamma e\to a e\gamma^\prime}/\sigma_\textrm{total}$, hence we will assume they are comparable.
Therefore, the dark photon and axion productions and their signals at the reactor experiments outweigh those at the FTNEs at most by the factor of $10^{6}$ per year, though backgrounds may also be far larger.
This comparison is also valid for other FTNEs such as LSND and MiniBooNE.

\begin{figure}[b]
\centering
\includegraphics[width=0.485\textwidth, trim={0 0 0 2.5mm}, clip]{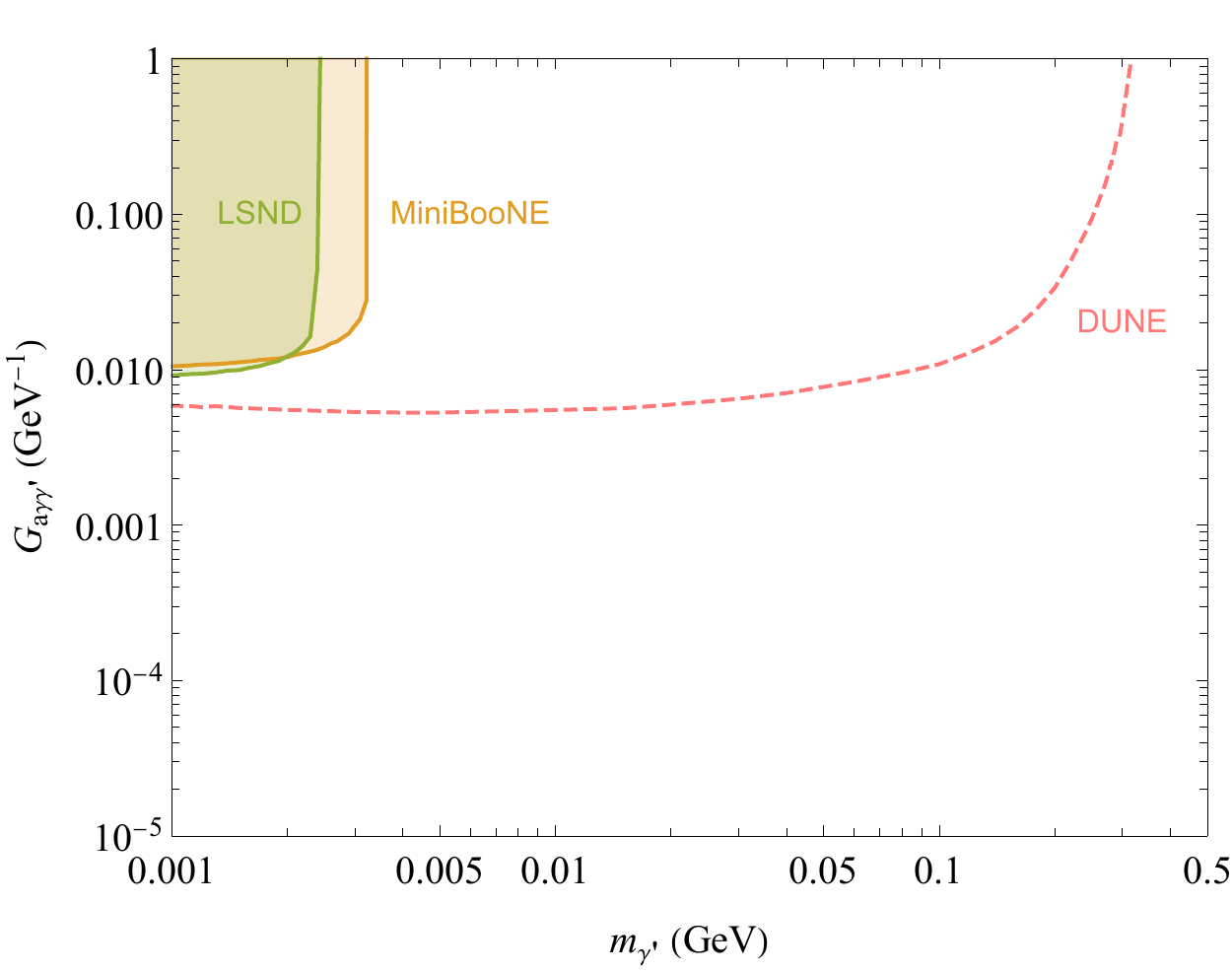}
\caption{Limits for fixed target neutrino experiments (FTNEs) with the inelastic scattering channels, $a/\gamma^\prime+e\rightarrow\gamma'/a+e$.
The LSND and MiniBooNE constraints were calculated with 90\% C.L. in Ref.~\cite{deNiverville:2018hrc}.
DUNE assumes 50\% efficiency and 5 years of data with 90\% C.L. and an on-axis DUNE detector position.
FTNEs have less ability to search for smaller couplings compared to the reactor experiments but are sensitive to larger masses.
}
\label{fig:ftne}
\end{figure}

We have also performed a projection of DUNE's sensitivity, drawing upon the background estimates of Ref.~\cite{Breitbach:2021gvv} and performing a counting experiment between Standard Model neutrino induced electron scattering events and those from the dark axion portal with a 90\% confidence level.
The projection assumes $5.5\times10^{21}$ POT (5 years of run) and 50\% efficiency with an on-axis DUNE detector position.
The inelastic scattering channels $a/\gamma^\prime+e\rightarrow\gamma'/a+e$ were considered as signals though most of the dark photons decay before reaching the detectors.
DUNE covers $m_{\gamma'}\lesssim0.3\,\textrm{GeV}$ and $G_{a\gamma\gamma'}\gtrsim0.005\,\textrm{GeV}^{-1}$ in the dark axion portal parameter space.
The LSND and MiniBooNE constraints were calculated with 90\% C.L. in Ref.~\cite{deNiverville:2018hrc}.
They have a coverage of $m_{\gamma'}\lesssim3\,\textrm{MeV}$ and $G_{a\gamma\gamma'}\gtrsim0.01\,\textrm{GeV}^{-1}$.
The limits of the FTNEs do not surpass the reactor experiments at smaller couplings.
However, future FTNEs can provide complementary sensitivity to reactor experiments in the larger masses.
The results are depicted in Fig.~\ref{fig:ftne}.

\begin{figure*}[t]
\centering
\subfigure[]{\includegraphics[width=0.4\textwidth]{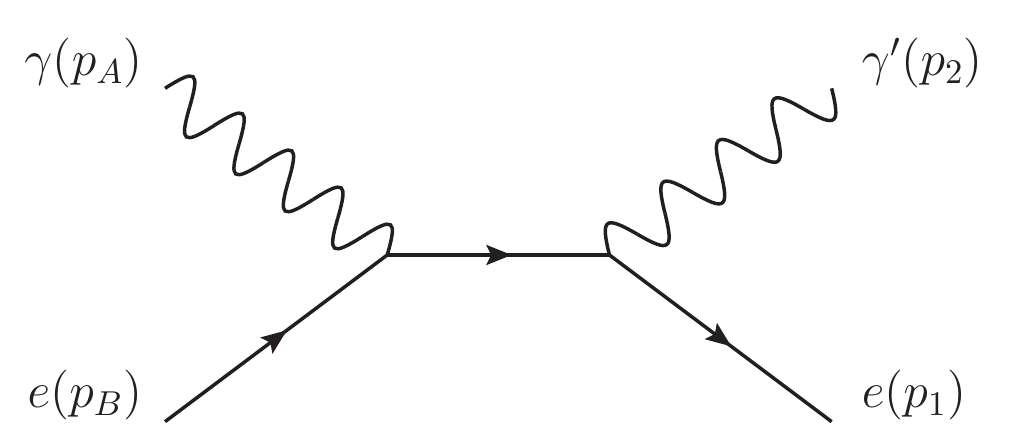}} ~~~
\subfigure[]{\includegraphics[width=0.3\textwidth]{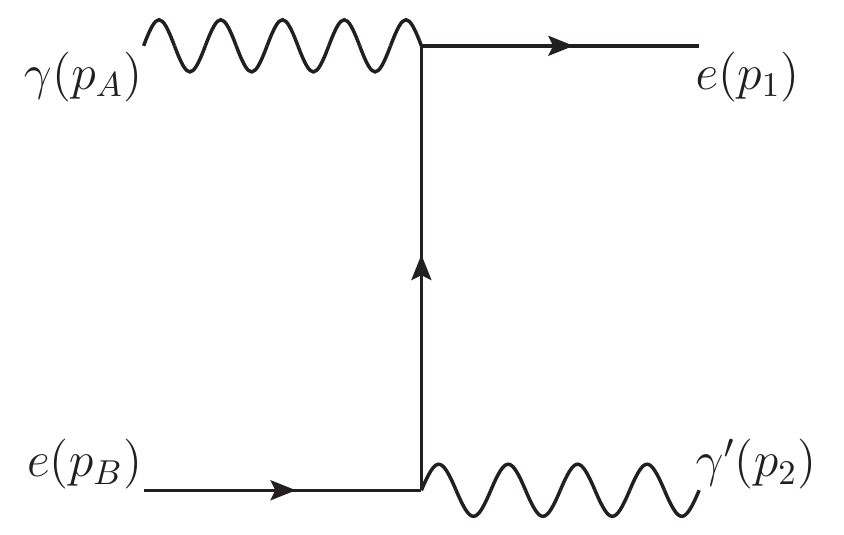}}
\caption{Feynman diagrams for the process $\gamma e \to \gamma^\prime e$ using the vector portal $\varepsilon$.
}
\label{fig:feynman_VP}
\end{figure*}

\subsection{Astrophysical constraints}
Studies on the astrophysical and cosmological constraints on $G_{a\gamma\gamma'}$ exist (e.g., see Refs.~\cite{Arias:2020tzl,Kalashev:2018bra,Daido:2018dmu}).
Perhaps most importantly, the stellar cooling condition provides a bound of $G_{a\gamma\gamma'}\lesssim10^{-9}\,\textrm{GeV}^{-1}$ when the dark photon mass is smaller than the plasma frequency in stars~\cite{Kalashev:2018bra,Daido:2018dmu}.

While the astrophysical constraints are very important, we do not study them here as our intention is to show the controlled lab experimental results in a rather model independent fashion.
The astrophysical constraints may be altered in the presence of other effects or new physics.
For instance, there are several models of axions or axionlike particles that evade the astrophysical constraints~\cite{Masso:2005ym,Khoury:2003aq,Brax:2007ak,Mohapatra:2006pv,Masso:2006gc,Dupays:2006dp,Jain:2005nh,Jaeckel:2006xm}.
Most of these models introduce a mechanism suppressing the stellar production of the axion; e.g., a model with an axion as a composite particle~\cite{Masso:2005ym}, a model with an axion as a chameleon-type field~\cite{Khoury:2003aq,Brax:2007ak}, a model with additional scalars~\cite{Mohapatra:2006pv} and a model with additional two dark photons~\cite{Masso:2006gc,Dupays:2006dp}.

Although the thorough discussion on the astrophysical bounds and developing mechanisms to avoid them for the dark axion portal will be important and interesting, it is beyond the scope of this paper and will be pursued in other works.

\section{The effect of the vector portal}
\label{sec:VP}

\begin{figure}[t]
 \centerline{\includegraphics[trim={0cm 0cm 0cm 0cm},clip,width=0.48\textwidth]{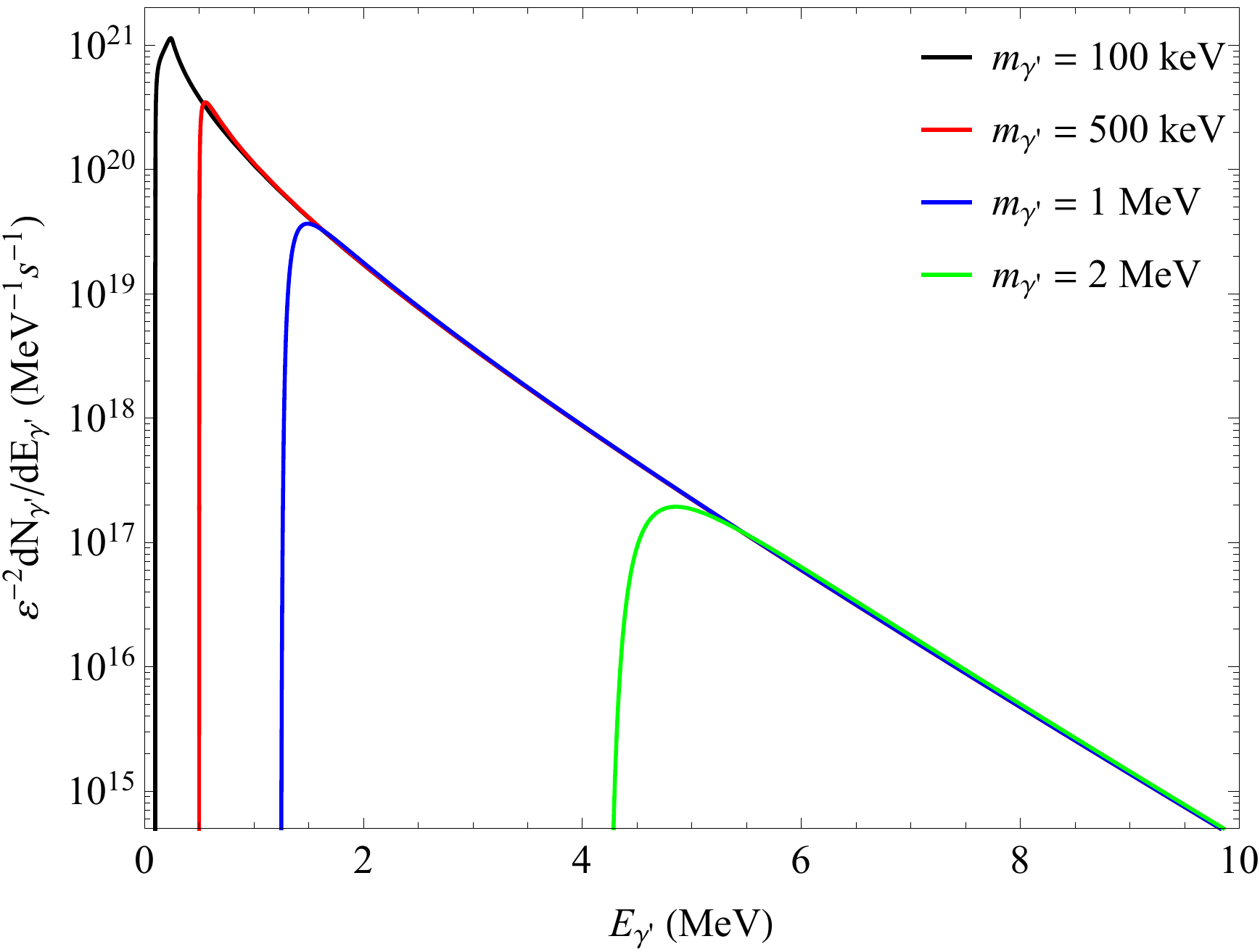}}
 \caption{Production spectrum of dark photons through the vector portal ($\varepsilon$) in a 1 GW nuclear reactor for several different dark photon masses. One can see that there are low energy kinematic cutoffs according to the dark photon mass.}
 \label{fig:production_vp}
\end{figure}

So far, we have assumed the vector portal to be absent but in the presence of the dark axion portal it is natural to also introduce the vector portal.
With a nonzero kinetic mixing, dark photons can be generated from the Compton-like process $e\gamma \to e \gamma^\prime$ (see Fig.~\ref{fig:feynman_VP}) as well as the dark axion portal.
The spectrum of $\gamma^\prime$ produced by reactor photons can be found with Eq.~\eqref{eq:dp_rate_dap} with $\sigma_{\gamma e\to a e\gamma^\prime}$ replaced by $\sigma_{e\gamma\to e\gamma^\prime}$.
As with the Compton scattering, both the $s$ and the $u$ channel diagram are considered, finding the expression with the help of FeynCalc
\begin{equation}
 |\mathcal{M}|^2 = \frac{32 \pi^2 \alpha^2 \varepsilon^2 (A + B)}{(m_e^2-s)^2 (m_e^2 - u)^2}
\end{equation}
where
\begin{align}
 A &=&&6 m_e^8 - 2 m_{\gamma'}^4 (m_e^2- s) (m_e^2 - u) \nonumber \\
& &&- s u (s^2 + u^2) + m_e^2 (s + u) (s^2 + 6 s u + u^2),\\
B &=&& - m_e^4 (3 s^2 + 14 s u + 3 u^2) \nonumber \\
& &&+  2 m_{\gamma'}^2(-4 m_e^2 s u + m_e^4 (s + u) + s u (s + u)). 
\end{align}
The $\gamma^\prime$ emission spectrum through the vector portal is shown in Fig.~\ref{fig:production_vp}.

On the detection side, the expected decay events can be calculated as before.
The dark photon production rate is negligibly small compared to the total number of photons in the reactor, hence the total observed signal can be found by simple addition of the individual signals from the dark axion portal and the vector portal.

In the presence of nonzero $\varepsilon$, if we consider $m_{\gamma^\prime}> 2 m_e$, then the leptonic decay $\gamma^\prime \to e^+ e^-$ is present;
\begin{equation}
 \Gamma_{\gamma^\prime\to e^+ e^-}=\frac{1}{3}\alpha \varepsilon^2 m_{\gamma'} \sqrt{1-\frac{4m_e^2}{m_{\gamma'}^2}} \left(1+\frac{2m_e^2}{m_{\gamma'}^2}\right),
\end{equation}
where $\alpha\equiv \frac{e^2}{4\pi}$ is the fine structure constant, though we will focus on the parameter space for which the decay through the dark axion portal is dominant. We neglect the $\gamma^\prime \to 3\gamma$ decay process as it is negligibly small~\cite{Pospelov:2008jk}.

\subsection{Contribution to the sensitivities}

For an MeV-scale dark photon, the kinetic mixing is bounded above on the order of $\varepsilon=10^{-8}$ by the electron beam dump experiments~\cite{Bjorken:2009mm,Andreas:2012mt,Jaeckel:2013ija}.
Here, we study how a nonzero kinetic mixing could affect the sensitivity of the $G_{a\gamma\gamma'}$ considering only the lab experiment bound; for the constraints from supernovae, see Refs.~\cite{Chang:2016ntp,DeRocco:2019njg}.
There is a possibility that the introduction of a new decay channel, $\gamma' \to a\gamma$, might relax the constraints on the kinetic mixing from the beam dump experiments.
Nevertheless, we found that the beam dump experiments constraints do not make a significant change for MeV-scale dark photons, and we keep $\varepsilon \lesssim 10^{-8}$.

\begin{figure}[t]
\centering
\includegraphics[width=0.48\textwidth]{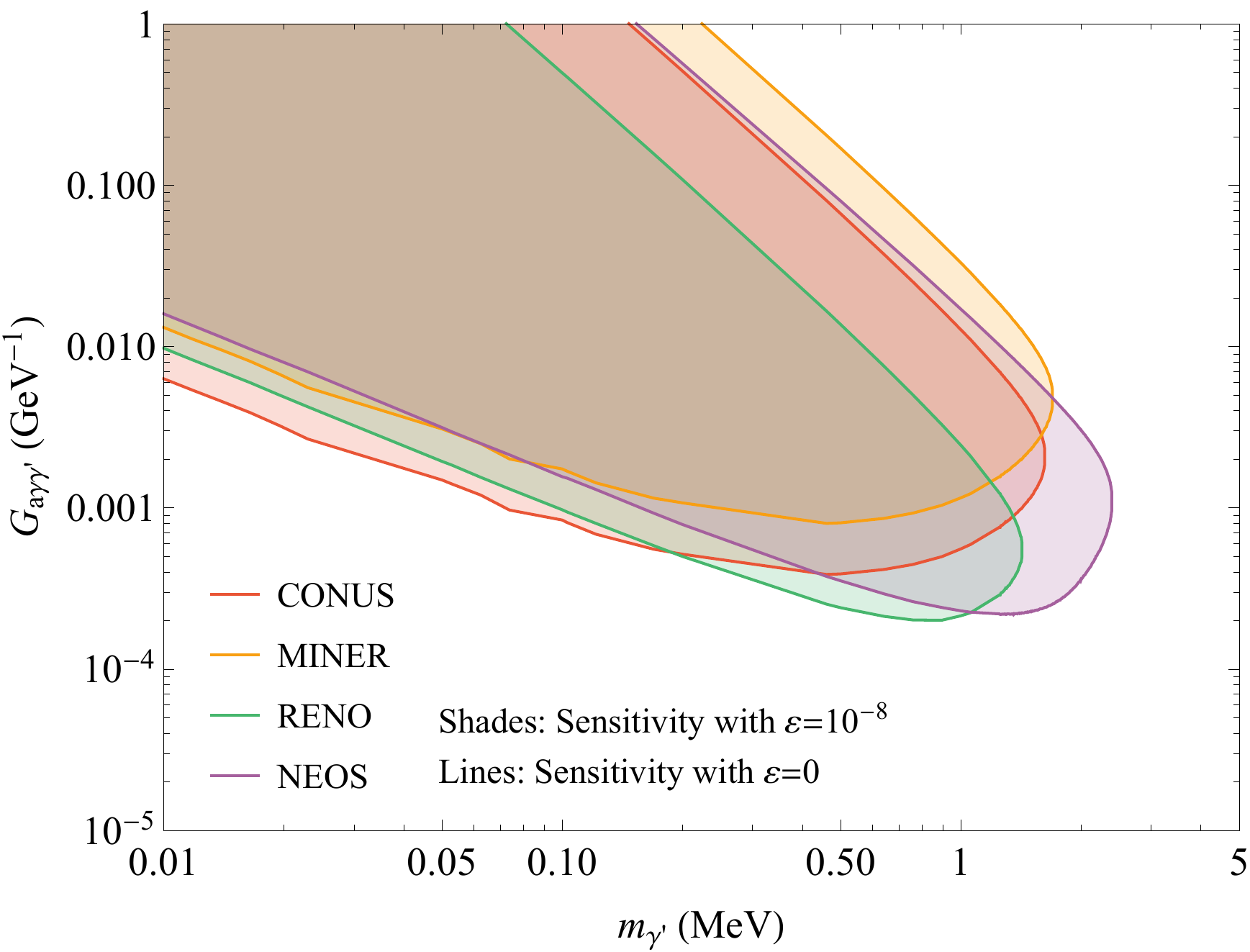}
\caption{
Expected sensitivities at CONUS, MINER, RENO, and NEOS in the presence of the dark axion portal and the vector portal.
Solid lines are from Fig.~\ref{fig:dap} ($\varepsilon=0$), while shaded regions are with nonzero kinetic mixing ($\varepsilon=10^{-8}$).
$\varepsilon=10^{-8}$ is chosen from the constraints from beam dump experiments.
Note that the shaded region is not a filling of the solid lines, but an individual calculation.
The difference in the sensitivities is nearly indistinguishable.
The production of the dark photon from the vector portal is suppressed by that from the dark axion portal because of small $\varepsilon$.
}
\label{fig:vp_eps-8}
\end{figure}

The $G_{a\gamma\gamma'}$ sensitivity with the level of $\varepsilon=10^{-8}$ is shown in Fig.~\ref{fig:vp_eps-8}.
The solid lines are the results of Fig.~\ref{fig:dap}, while the shaded region is the sensitivity under the presence of the kinetic mixing.
Note that the shaded region is the result for a separate calculation, not just a padding of the solid lines.
We can see the effect of the vector portal is negligible in the case of $\varepsilon=10^{-8}$.
The production of the dark photon from the dark axion portal is dominant over that from the vector portal as $\varepsilon$ is constrained to be small.
This becomes apparent with a simple calculation using the production figures of the dark axion portal (Fig.~\ref{fig:production_dap}) and vector portal (Fig.~\ref{fig:production_vp}).
For MeV-scale dark photons, the region of interest in the dark axion portal coupling is on the order of $G_{a\gamma\gamma'}=10^{-4}\,\textrm{GeV}^{-1}$ (see Fig.~\ref{fig:dap}).
Then the dark photon production from the dark axion portal is on the order of $10^{4}\,\textrm{MeV}^{-1}s^{-1}$ for $m_{\gamma'}=1\,\textrm{MeV}$.
On the other hand, the dark photon production from the vector portal is on the order of $10^{3}\,\textrm{MeV}^{-1}s^{-1}$ in the case of $\varepsilon=10^{-8}$, which is buried in the signals from the dark axion portal.

\subsection{Implications of additional light dark particles}
When there are light dark sector particles other than axions and dark photons, the dark photon may decay to these invisible particles, and the invisible decay can weaken the constraints on kinetic mixing from beam dump experiments.
The constraints, in this case, depend on $m_\chi$ and $\alpha_D$, where $m_\chi$ is the dark matter mass and $\alpha_D$ is the dark fine structure constant $\alpha_D\equiv\frac{{e'}^2}{4\pi}{Q'_\chi}^2$.
As an illustration, for sub-MeV dark matter, the beam dump experiments give $\varepsilon^2\sqrt{\alpha_D}\lesssim 10^{-10}$ and the $BABAR$ experiment gives $\varepsilon\lesssim 10^{-3}$~\cite{Izaguirre:2015yja,Akesson:2018vlm}.
We used these constraints to study the impact of the additional light dark particles on the sensitivity of reactor experiments to $G_{a\gamma\gamma'}$, though the dark axion portal's exact effect on these constraints should be further investigated.

We found that the reactor experiments studied lose sensitivity for $m_{\gamma'}\geq2m_\chi$ even when a larger kinetic mixing is present, e.g. $\varepsilon\simeq10^{-4}$ with $\alpha_D=10^{-8}$.
This loss of sensitivity is because (1) the $\gamma' \to a\gamma$ branching ratio is suppressed by introducing the invisible $\gamma' \to \chi\bar\chi$ channel and (2) the decay length becomes very short due to the increased decay width.
Even smaller $\alpha_D$ could relieve the dominance of the invisible decay, but it is forbidden by the supernova constraint together with the beam dump constraints.
The supernova constraint requires that $\varepsilon^2\alpha_D\gtrsim10^{-14}$ for $m_{\gamma'}\sim m_\chi\sim 10\, \textrm{MeV}$~\cite{Izaguirre:2014bca,Izaguirre:2015yja}.
Although the mass range of $m_{\gamma'}$ differs from our mass range of interest by one order of magnitude, it is apparent that extremely small $\alpha_D$ should be avoided.
It would be beneficial to consider the scattering of the light dark particle as a signal as well.

\begin{figure*}[tb]
\centering
\includegraphics[trim=0 0 0 12.5,clip,width=0.6\textwidth]{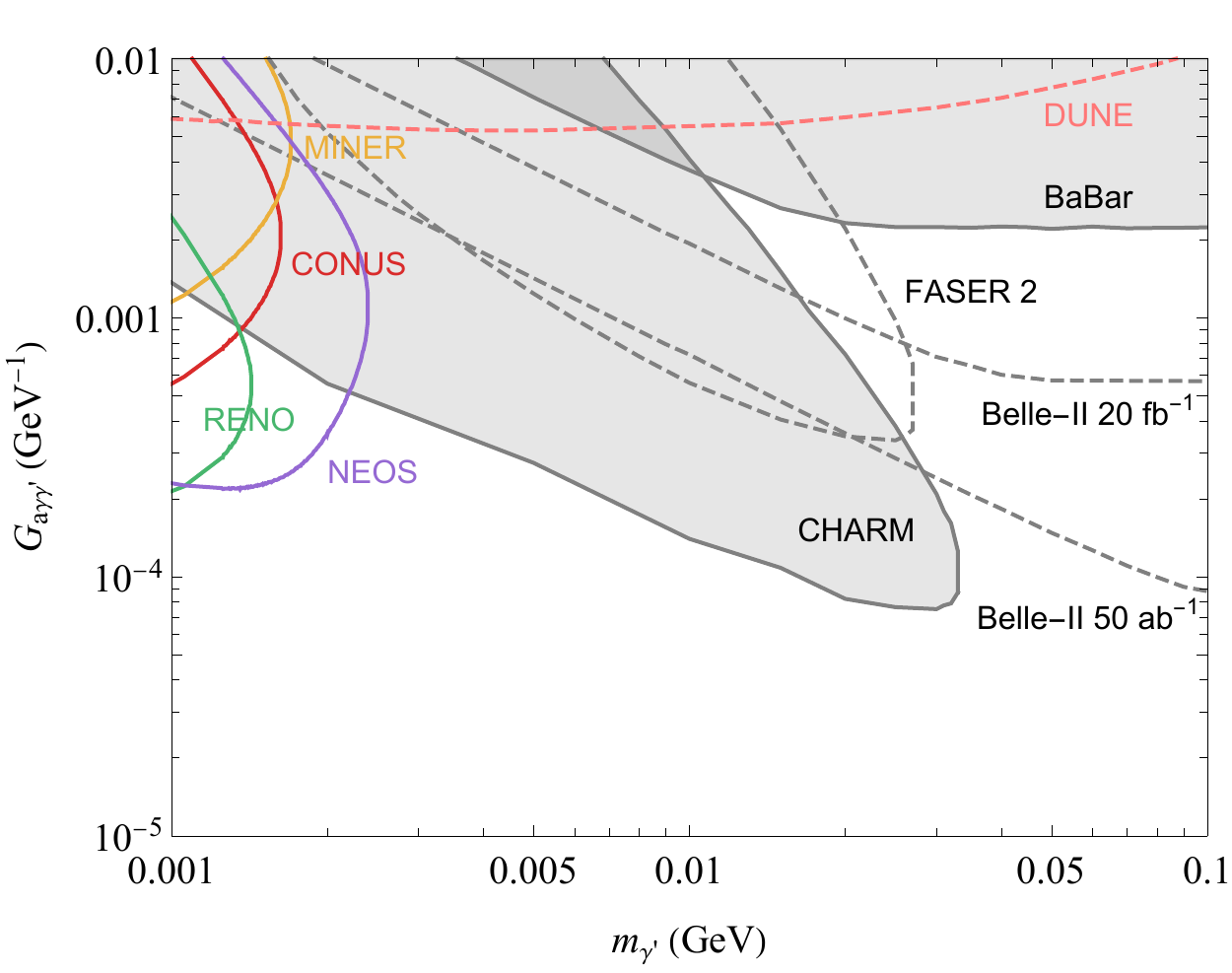}
\caption{Expected sensitivities for RENO, NEOS, CONUS, MINER from Fig.~\ref{fig:dap} and DUNE from Fig.~\ref{fig:ftne} where $\varepsilon=0$ with other limits~\cite{deNiverville:2018hrc,deNiverville:2019xsx}.
The shaded regions are the experimental constraints while the dashed or colored lines are the expected sensitivities.
The axion mass is assumed to be negligibly small.
Note that reactor experiments can access the parameter space especially the low mass and low coupling region where other experiments are insensitive.
}
\label{fig:dap_all}
\end{figure*}

\section{Summary and Discussion}
\label{sec:summary}

Reactors have been used for the search of light particles such as neutrinos, axions and dark photons.
Through the chain reaction in the reactor core, a huge flux of photons and neutrinos is generated, and it was through this flux that neutrinos were first observed, and a precise measurement of the neutrino oscillation parameter was made.
The enormous available flux renders these experiments sensitive to very weakly interacting particles, making them ideal laboratories in which to search for MeV-scale dark states. Sterile neutrino searches and coherent elastic neutrino-nucleus scattering measurements utilizing reactors are also active. Various axion models, axionlike particles and dark photons have been tested and constrained through reactor experiments.

The dark axion portal arises as an axion-photon-dark photon vertex in the presence of the axion and dark photon.
Because of the dark gauge coupling of the dark photon with the exotic fermions in the anomaly triangle, the dark axion portal is an independent portal from the vector portal and axion portal.
Though dark photon searches typically rely on the vector portal (the kinetic mixing between photons and dark photons), the dark axion portal can introduce new production and decay channels.

We investigated the possibility of detecting dark photons using the dark axion portal at reactor neutrino experiments.
We considered four experiments out of many nuclear reactor experiments; MINER and CONUS were designed to measure the coherent elastic neutrino-nucleus scattering, while RENO and NEOS aimed to measure neutrino oscillation parameters and sterile neutrinos.
We show the expected sensitivities in the dark photon mass ($m_{\gamma'}$) and dark axion portal coupling ($G_{a\gamma\gamma'}$) parameter space in Fig.~\ref{fig:dap_all} for the kinetic mixing $\varepsilon=0$ case.
A highly conservative background analysis was performed, but it could be further improved to increase coverage of the parameter space.

Figure~\ref{fig:dap_all} also shows the constraints and sensitivities from other experiments for $m_{\gamma'} \geq 1\,\mathrm{MeV}$~\cite{deNiverville:2018hrc,deNiverville:2019xsx}.
Here, we consider only the lab and reactor experiments; for the constraints in the region of $m_{\gamma'}\lesssim1\,\textrm{MeV}$, see Ref.~\cite{Arias:2020tzl}, for instance.
The shaded regions/colored and dashed curves represent the experimental constraints/expected sensitivities.
It is worth noting that the coverage of reactor experiments in the parameter space is complementary to those of other experiments.
One possible optimal design might be a RENO sized detector at the NEOS position; the installation of a decay volume does not provide much benefit due to the wide spread of the decay products, and resulting poor acceptance in the instrumented region.
Furthermore, another type of neutrino experiment, the fixed target neutrino experiment, possesses complementary sensitivity to the reactor neutrino experiments, reaching larger masses but without the ability to probe smaller couplings.

We also studied if the vector portal could alter the results.
The beam dump constraints on $\varepsilon$ do not change even with the new decay channel, $\gamma' \to a\gamma$, for the MeV-scale dark photons.
Accounting for the upper bound of $\varepsilon=10^{-8}$ from beam dump experiments, there was no visible enhancement in the sensitivity when the vector portal was included in the analysis.
The existence of additional light dark matter states could weaken the constraints from beam dump experiments, for the dark photon can also decay into a dark matter pair, $\gamma'\to\chi\bar\chi$, through the invisible decay channel.
However, even with a larger kinetic mixing the coverage in the parameter space was reduced as the invisible decay dominates over the $\gamma' \to a\gamma$ decay and the decay length of the dark photon becomes too short.

In short, we investigated the effect of the dark axion portal with and without the vector portal in reactor experiments for the first time.
As our study shows, $m_{\gamma'}\lesssim\textrm{few\,MeV}$ and $G_{a\gamma\gamma'}\gtrsim10^{-4} \, \textrm{GeV}^{-1}$ can be covered with the currently running reactor experiments.
These low mass, low coupling regions were not covered by other experiments but can be well probed with the experiments using high power nuclear reactors.
We took a rather conservative approach, and the coverage in the parameter space might be significantly improved to larger masses and lower couplings if we include more channels and perform detailed background analysis.
An immediate analysis of the data of the existing reactor experiments is well motivated.

\onecolumngrid

\section*{Acknowledgements}
This work was supported in part by Los Alamos National Laboratory under the LDRD program and the National Research Foundation of Korea (No. NRF-2017R1E1A1A01072736, No. NRF-2019R1A6A1A10073887).
H.L. thanks the Erwin Schr\"odinger International Institute and TRIUMF for hospitality while part of this work was completed.
We thank Y.D. Kim, Y.J. Ko, Y. Oh, S. Seo, and J. Yoo for helpful discussions about the reactor experiments.

\twocolumngrid

\bibliography{./main}{}

\end{document}